\documentclass[conference,dvipdfmx]{IEEEtran}
\usepackage{amssymb,stmaryrd,amsmath,amsfonts,rotating}
\usepackage{amsmath,amsthm,amssymb,cases}
\usepackage{amsmath,cases}
\usepackage{amsthm}
\usepackage{color}
\usepackage{amssymb}
\usepackage{bm}
\usepackage{mathtools}
\mathtoolsset{showonlyrefs}

\newcommand{\R}{\mathbb{R}}

\newcommand{\dg}{d_g}
\newcommand{\dl}{d_l}

\newcommand{\dr}{d_r}
\theoremstyle{definition}

\newtheorem{theorem}{Theorem}

\newtheorem{definition}{Definition}
\newtheorem{example}{Example}
\newtheorem{lemma}{Lemma}

\newcommand{\e}{\epsilon}

\newcommand{\x}{\bm{x}}
\newcommand{\y}{\bm{y}}

\newcommand{\D}{\bm{D}}
\newcommand{\f}{\bm{f}}
\newcommand{\g}{\bm{g}}

\newcommand{\Xc}{\mathcal{X}}

\newcommand{\Fc}{\mathcal{F}}

\newcommand{\Z}{\mathbb{Z}}

\ifCLASSINFOpdf
\else
\fi
\begin{document}
%
\title{ Spatially-Coupled MacKay-Neal Codes \\ Universally Achieve the Symmetric Information Rate of Arbitrary Generalized Erasure Channels with Memory}
\author{
\IEEEauthorblockN{Masaru Fukushima, Takuya Okazaki and Kenta Kasai}
\IEEEauthorblockA{Department of Communications and 
Computer Engineering,\\
 Tokyo Institution of Technology\\
Email: \{fukushima,osa,kenta\}@comm.ce.titech.ac.jp}

}
\maketitle

\begin{abstract}
This paper investigates the belief propagation decoding of spatially-coupled MacKay-Neal (SC-MN) codes over erasure channels with memory. 
We show that SC-MN codes with bounded degree universally achieve the symmetric information rate (SIR) of arbitrary erasure channels with memory.
We mean by universality  the following sense: the sender does not need to know the whole channel statistics but needs to know only the SIR, while
the receiver estimates the transmitted codewords from channel statistics and received words. 
The proof is based on potential function. 
\end{abstract}
\IEEEpeerreviewmaketitle

\section{Introduction}
Felstr\"{o}m and Zigangirov introduced  convolutional LDPC codes \cite{zigangirov99}. 
Later the codes are called spatially-coupled (SC) codes. 
Lentmaier et al. confirmed that regular SC LDPC codes achieve  excellent BP decoding \cite{lentmaier_II}. 
Kudekar et al. proved that SC codes achieve MAP threshold by BP decoding on the binary erasure channel (BEC) \cite{5695130} 
and binary memoryless symmetric channels \cite{6589171}.

In \cite{journals/ieicet/TakeuchiTK12}, Takeuchi et al. introduced potential function to understand how threshold saturation phenomenon happen. 
With some modifications, Yedla et al. proved threshold saturation of spatially-coupled LDPC codes over BEC in \cite{simple_proof_vector_turbo,6887298}. 

Kasai et al.~introduced SC MacKay-Neal (MN) codes, and showed that these codes with finite maximum degree achieve capacity of BEC by numerical experiment \cite{HSU_MN_IEICE}. 
Obata et al. \cite{ISIT_OJKP} and Okazaki et al. \cite{6874884} proved respectively, $(\dl,\dr=2,\dr=2)$ SC-MN codes 
and $(\dl,\dr=3,\dg=3)$ SC-MN codes achieve the capacity of BEC 
for any $\dl>\dr$, where each of $\dl,\dr,\dg$ describes the degree of nodes in the Tanner graph of the codes.

Reliable transmissions over channels with memory are practically important, e.g., magnetic recording with ISI and Rayleigh fading channel with memory \cite{910595}. 
Pfister and Siegel proved that carefully designed irregular LDPC codes can achieve the symmetric information rate (SIR) of dicode erasure channels (DEC) under joint iterative decoding \cite{1003830}. 
In \cite{KuKaDEC} and \cite{SekidoPR2_SITA}, it was empirically observed that the BP threshold of  spatially-couple regular codes achieve the SIR of DEC and PR2 channels, respectively, by increasing node degree. 

In this paper, we show that SC-MN codes with bounded degree universally achieve the 
symmetric information rate of wide variety of erasure channels with memory.
We mean by universality  the following sense: the sender does not need to know the whole channel statistics 
$p(\underline{y}|\underline{x})$ but needs to know only the SIR, while
the receiver estimates the transmitted codewords from channel statistics and received words. 
The proof is based on the powerful potential function method \cite{journals/ieicet/TakeuchiTK12,simple_proof_vector_turbo,6887298}. 
\section{Background}
\subsection{Generalized Erasure Channel (GEC)}
Denote a channel input and output as $\underline{x}=(x_1,\dotsc,x_n)\in \{0,1\}^n$ and $\underline{y}=(y_1,\dotsc,y_n)\in \mathcal{Y}^n$, respectively. 
We assume that $\underline{x}\in \{0,1\}^n$ is uniformly  distributed. 
Let us  assume that by introducing some appropriate state nodes $\underline{\sigma}=(\sigma_1,\dotsc,\sigma_n)$  such that 
\begin{align}
 p(\underline{x}|\underline{y})=\sum_{\underline{\sigma}}p(\underline{x},\underline{\sigma}|\underline{y}),
\end{align} 
we can factorize $p(\underline{x},\underline{\sigma}|\underline{y})$ so that its factor graph is  a tree. 
By Bayes rule, we have $p(\underline{x},\underline{\sigma}|\underline{y})\propto p(\underline{y}|\underline{\sigma},\underline{x})p(\underline{\sigma}|\underline{x})p(\underline{x})$. 
Consider the APP detector implemented by sum-product algorithm to calculate 
\begin{align}
 \bigl(p(x_j=0|\underline{y}),p(x_j=1|\underline{y})\bigr)=:\bigl(\mu_j(0),\mu_j(1)\bigr)
\end{align}
for $j= 1,2,\dotsc,n.$
We say the channel $p(\underline{y}|\underline{x})$ is a generalized erasure channel (GEC)
if $(\mu_j(0),\mu_j(1))$ is one of $\{(1,0), (0,1), (1/2,1/2)\}$. 
The corresponding LLR values are $+\infty$, $-\infty$ and $0$. Some authors use $0$, $1$ and $?$, instead. 

\subsection{LDPC codes over GEC}
Next, consider $\underline{x}$ is uniformly distributed in an LDPC code $C$.  In precise, 
\begin{align}
 p(\underline{x})=
\begin{cases}
 1/\#C&(\underline{x}\in C)\\
 0&(\underline{x}\notin C)
\end{cases} 
\end{align}
Let us denote the factor graph of $p(\underline{y}|\underline{\sigma},\underline{x})p(\underline{\sigma}|\underline{x})p(\underline{x})$ by $G$. 
We divide $G$ into two subgraphs. 
One is the factor graph  of $p(\underline{y}|\underline{\sigma},\underline{x})p(\underline{\sigma}|\underline{x})$ and the other is the factor graph of $p(\underline{x})$. 
They are corresponding to   APP detector and LDPC decoder, respectively. 
Since the message from APP detector to the bit nodes is one of $\mathcal{M}$, 
the messages used in the LDPC decoder also takes value in $\mathcal{M}$. 

Consider the density evolution of the BP decoding on $G$ in the limit of code length.
We define $\phi(x;\e)$ as a function which maps 
the erasure probability of messages from LDPC decoder to APP detector via bit nodes $x$, to 
the erasure probability of messages from APP detector to from LDPC decoder $\phi(x;\e)$, 
where $\epsilon$ is a parameter which defines the channel. 
From the definition, it follows that  $\phi(x;\e)$ is non-decreasing in $x\in[0,1]$. 
In this paper, we further assume that $\phi(x;\e)$ is twice continuously differentiable with $x$ and strictly increasing with $\epsilon\in[0,1]$. 
In this setting, the density evolution for $(\dl,\dr)$ regular LDPC codes over GEC with $\phi(x;\e)$ is written as follows. 
\begin{align}
&  x^{(0)}=1, \\
&  x^{(t+1)}=\phi\bigl((1-(1-x^{(t)})^{\dr-1})^{\dl};\e\bigr)\\
&\hspace{2cm}\cdot(1-(1-x^{(t)})^{\dr-1})^{\dl-1}, 
\end{align}
where $x^{(t)}$ is the erasure probability of messages from bit nodes to parity-check nodes at the $t$-th decoding round. 
\subsection{Symmetric Information Rate}
SIR $I(\e)$ is the mutual information is defined as follows
\begin{align}
  I(\e):=\lim_{n\to\infty}\frac{1}{n}I(\underline{X};\underline{Y}(\epsilon)),
\end{align}
under the existence of limit, where capital letters represent random variables.
In \cite{4444763,1347354}, it is shown that the SIR is calculated via $\phi(x;\e)$ as follows. 
\begin{align}
  I(\e)=1-\int_0^1 \phi(x;\e)dx=1-\Phi(1;\e),
\end{align}
where $\Phi(x;\e) := \int_0^x\phi(x';\e)dx'$.
Let $R$ be the coding rate.
Define SIR limit as $\epsilon$ such that $I(\epsilon)=R$, and denote it by 
$\epsilon^{\mathrm{SIR}}(R)$.
Uniqueness of $\epsilon^{\mathrm{SIR}}(R)$ is again due to the assumption that $\phi(x;\epsilon)$ is increasing in $\e$. 
\subsection{MacKay-Neal Codes}
$(\dl,\dr,\dg)$ MN codes are multi-edge type (MET) LDPC used over GEC with $\phi(x;\e)$ codes \cite{mct} defined by pair of multi-variables degree distributions $(\mu,\nu)$ listed below.
\begin{align}
 \label{MN:dist}
 \nu(\x;\Phi(x;\e)) &= \frac{\dr}{\dl}x_1^{\dl}+\Phi(x_{2}^{\dg};\e), \quad\mu(\x) =x_1^{\dr}  x_2^{\dg}.
\end{align}
Here, we slightly extended the definition of degree distribution in such a way that 
the bits corresponding to the term $\Phi(x_{2}^{\dg};\e)$ are transmitted through the GEC with $\phi_\e$. 
In the case of BEC($\epsilon$), $\Phi(x,\epsilon)=\epsilon x$. 
We define the erasure probability message sent from bit nodes along edges of type $j$  at the $t$-th  decoding round by $x_j^{(t)}$. 
The recursion of density evolution of MET-LDPC codes on BEC is given by
   \begin{align}
x_{j}^{(t)}=\frac{\nu_{j}(\y^{(t)};\Phi(x;\e))}{\nu_{j}(\bm{1};\mathrm{id}_\R)}, \quad y_{j}^{(t+1)}  &=1-\frac{\mu_{j}(\bm{1}-\x^{(t)})}{\mu_j(\bm{1})},
   \end{align}
 where $\nu_j(\x;\Phi(x;\e)):=\frac{\partial}{\partial x_j}\nu(\x;\Phi(x;\e))$, $\mu_j(\x):=\frac{\partial}{\partial x_j}\mu(\x)$ and $\mathrm{id}_\R$ is the identity function  $\mathrm{id}_\R(x)=x$. 

Then, the density evolution of $(\dl,\dr,\dg)$ MN codes is
 \begin{align}
  \x^{(0)}&=\bold{1}, \quad  \x^{(t+1)} = \f(\g(\x^{(t)});\e),\label{eq:mnde}
 \end{align}
where
\begin{align}
  \f(\x;\e):=& \bigl(x_{1}^{\dl-1}, \phi(x_{2}^{\dg};\e)\bigr)\label{eq:mndef}\\
 \g(\x):=&\bigl(1-(1-x_{1})^{\dr-1} (1-x_{2})^{\dg},  \\
&\phantom{\bigl(} 1-(1-x_{1})^{\dr}(1-x_{2})^{\dg -1}  \bigr).\label{eq:mndeg}
\end{align}
\subsection{Spatially-Coupled MacKay-Neal Codes}
 SC-MN codes of chain length $L$ and of coupling width $w$ are defined by the Tanner graph constructed by the following process. 
First,  at each section $i\in \Z$, place $rM/l$ bit nodes of type 1 and $M$ bits nodes of type 2. 
Bit nodes of type 1 and 2 are of degree $\dl$ and $\dg$, respectively. 
Next,  at each section $i \in \Z$, place $M$ check nodes of degree $\dr+\dg$. 
Then, connect edges uniformly at random so that 
bit nodes of type 1 at section $i$ are connected with check nodes at each section $i\in [i,\dotsc, i+w-1]$ with $\dr M/w$ edges, and
bit nodes of type 2 at section $i$ are connected with check nodes at each section $i\in [i,\dotsc, i+w-1]$ with $\dg M/w$ edges. 
Bits at section $i\notin [0,L-1])$ are shortened. 
Bits of type 1 and 2 at section $i\in[0,L-1]$ are punctured and transmitted, respectively. 
The rate of SC-MN codes $R^{MN}(\dl,\dr,\dg,L,w)$ is given by
\begin{align}
&R^{\rm{MN}}(\dl,\dr,\dg,L,w)\\
&=\frac{\dr}{\dl}+\frac{1+w-2\sum_{i=0}^{w}(1-(\frac{i}{w})^{\dr+\dg})}{L} \\
&=\frac{\dr}{\dl}\quad (L\to\infty).\label{010653_25Jan15}
\end{align}
\subsection{Vector Admissible System and Potential Function}
In this section, we define vector admissible systems and potential functions. 
\begin{definition}
\label{def:vas}
{ 
Define $\Xc\triangleq[0,1]^d$, and   $F : \Xc \times [0, 1]\to \mathbb{R}$ and $G : \Xc \to \mathbb{R}$ as functional satisfying $G(\bm{0})=0$. Let $\D$ be a $d \times d$ positive diagonal matrix. Consider a general recursion defined by
  \begin{equation}
    \bm{x}^{(\ell+1)}=\bm{f}(\bm{g}(\bm{x}^{(\ell)});\e) \label{vas}
  \end{equation}
where $\f : \Xc \times [0,1] \to \Xc$ and $\g : \Xc \to \Xc$ are defined by  $F'(\x;\e) = \f(\x;\e) \D$ and $G'(\x) = \g(\x) \D$, 
where $F'(\x;\e)\triangleq (\frac{\partial F(\x)}{\partial x_1},\dotsc,\frac{\partial F(\x)}{\partial x_n})$. 
Then the pair $(\f,\g)$ defines a vector admissible system if
 \begin{enumerate}
\item $\f,\g$ are twice continuously differentiable,
\item $\f(\x;\e)$ and $\g(\x)$ are non-decreasing in  $\x$ and $\e$ with respect to $\preceq$ 
\footnote{We say $\x \preceq \y$ if $x_i \leq y_i$ for all $1 \le i \le d$},
\item $\f(\g(\bm{0});\e)=\bm{0}$ and $F(\g(\bm{0});\e)=0$.
\end{enumerate}
We say  $\x$ is a fixed point if $\x=\f(\g(\x);\e)$.
}
\end{definition}

\begin{definition}[{\cite[Def.~2]{simple_proof_BEC_itw}}]
  \label{def:potential}
 We define the potential function $U(\x;\e)$ of a vector admissible system $(\f ,\g)$  by
 \begin{align}
    \label{potential}
  U(\x ;\e) \triangleq \g(\x)\D\x^T-G(\x)-F(\g(\x);\e).
 \end{align}
\end{definition}

\begin{definition}[{\cite[Def. 7]{simple_proof_BEC_itw}}]
\label{def:PotentialThreshold}
{
 Let $\Fc(\e) \triangleq \{\x \in \Xc\setminus\{\bm{0}\} \mid
\x=\f(\g(\x);\e)\}$ be a set of non-zero fixed points for $\e \in [0,1]$. The potential threshold $\e^{*}$ is defined by
  \begin{equation*}
  \label{Ue}
    \e^{*} \triangleq \sup \{ \e \in [0,1] \mid \min\nolimits_{\x \in \mathcal{F}(\e)} U(\x;\e) > 0 \}.
  \end{equation*}
Let $\e_s^*$ be threshold of uncoupled system defined in \cite[Def. 6]{simple_proof_BEC_itw}. 
For $\e$ such that $\e_s^* < \e < \e^*$, we define energy gap $\Delta E(\e)$ as
  $$\Delta E(\e) \triangleq \max_{\e'\in [\e,1]} \inf_{ \x\in \mathcal{F}(\e')} U(\x;\e').$$
 }
\end{definition}

\begin{definition}[{\cite[Def.~9]{simple_proof_BEC_itw}}]
\label{def:SpatiallyCoupled}
For  a  vector admissible system $(\f,\g)$, we define the SC system of chain length $L$ and coupling width $w$ as 
 \begin{align}
\label{eq:4}
  \x^{(t+1)}_i &=
  \frac{1}{w}\sum_{k=0}^{w-1}\f\Bigg(\frac{1}{w}\sum_{j=0}^{w-1}\g(\x^{(t)}_{i+j-k});\e_{i-k}\Bigg),  \\
 \e_{i} &= 
\begin{cases}
\e,& i\in    \{0,\dotsc,L-1\},\\
 0,      & i\notin \{0,\dotsc,L-1\}.
\end{cases}
\end{align}
\end{definition}
If we define $(\f,\g)$ as the density evolution for $(\dl,\dr,\dg)$ MN codes in \eqref{eq:mndef} and \eqref{eq:mndeg}, the SC system gives
the density evolution of SC-MN codes with chain length $L$ and coupling width $w$. 

Next theorem asserts that if $\e < \e^*$ then fixed points of SC vector system converge towards $\bold{0}$ for sufficiently large $w$.
\begin{theorem}[{\cite[Thm.~1]{simple_proof_BEC_itw}}]
\label{theorem:sc_theorem}
Consider the constant $K_{\f,\g}$ defined in \rm{\cite[Lem.~11]{simple_proof_BEC_itw}}. 
This constant value depends only on $(\f,\g)$. 
If $\e < \e^*$ and $w > (d K_{\f,\g})/( 2\Delta E(\e))$, then the SC system of $(\f, \g)$ with chain length $L$ and coupling width $w$ has a unique fixed point $\bold{0}$.
\end{theorem}

It can be seen that the density evolution  $(\f,\g)$ of $(\dl,\dr,\dg)$ MN codes  over GEC($\phi(x;\e)$) is a vector admissible system by choosing
 $F\bigl(\x;\e\bigr), G(x)$ and $\D$ as below, since this system $(\f,\g)$ satisfies the condition in Definition \ref{def:vas}. 
\begin{align*}
  F(\x;\e) &= \frac{\dr}{\dl}x_{1}^{\dl}+\Phi(x_{2}^{\dg};\e),  \\
  G(\x) &= \dr x_1+\dg x_2+(1-x_1)^{\dr}(1-x_2)^{\dg}-1, \\ 
  \bm{D} &= \left(
  \begin{array}{cc}
    \dr & 0 \\
    0 & \dg
    \end{array}
  \right) .
\end{align*}
From Definition \ref{def:potential}, the potential function $U(x_1,x_2,\e)$ of $(\dl,\dr,\dg)$ MN codes is given by
\begin{align}
&U(x_1,x_2;\e)) \label{173605_11Jan15} \\
&= 1-\Phi(\{1-(1-x_1)^{\dr}(1-x_2)^{\dg-1}\}^{\dg};{\e})  \\
&\quad- \frac{\dr}{\dl}\{1-(1-x_1)^{\dr-1}(1-x_2)^{\dg}\}^{\dl}  \\
&\quad- (1-x_1)^{\dr}(1-x_2)^{\dg}  \cdot \Bigl(1+\frac{\dr x_1}{1-x_1}+\frac{\dg x_2}{1-x_2}\Bigr).
\end{align}

\section{Proof of Achieving SIR}
In this section, we will prove that $(\dl,\dr=2,\dg=2)$ and $(\dl,\dr=3,\dg=3)$ SC-MN codes, 
for any $\dl>\dr$,  achieve the SIR of any GEC.
First, we calculate the potential threshold $\e^*(\dl,\dr,\dg)$ of MN codes, and show that the potential threshold equals to the SIR limit $\e^\mathrm{SIR}(\dr/\dr)$. 
Then we apply Theorem \ref{theorem:sc_theorem} which  proves that density evolution of SC-MN code has a unique fixed point $\bold{0}$ for $\e$ smaller than SIR limit $\e^\mathrm{SIR}(\dr/\dr)$.

\subsection{Potential Function at Trivial Fixed Point}
Recall the definition of potential threshold $\e^*(\dl,\dr,\dg)$ in Definition \ref{def:PotentialThreshold}. 
We need to investigate the structure of  $\mathcal{F}(\epsilon)$ to calculate the potential threshold $\e^*(\dl,\dr,\dg)$. 
The density evolution \eqref{vas} of  $(\dl,\dr,\dg)$ MN codes at fixed point $(x_1,x_2;\phi(x;\e))$ can be rewritten as
\begin{align}
x_{1}=&(1-(1-x_{1})^{\dr-1}(1-x_{2})^{\dg})^{\dl-1},\label{145556_10Jan15} \\
x_{2}=&\phi(\{1-(1-x_{1})^{\dr}(1-x_{2})^{\dg-1}\}^{\dg};{\e})\label{145604_10Jan15}\\
&\cdot\{1-(1-x_{1})^{\dr}(1-x_{2})^{\dg-1}\}^{\dg-1}. \label{vasx2}
\end{align}

First, observe that $(x_1=1,x_2=\phi(1;\epsilon))\in\mathcal{F}(\epsilon)$ for $\epsilon\in [0,1]$.
We call these fixed points trivial. 
From \eqref{173605_11Jan15} and the definition of SIR limit, the next lemma asserts that 
the sign of $U\bigl(1,\phi(1;\epsilon);\e\bigr)$ changes at the SIR limit $\e^{\mathrm{SIR}}\bigl(\textstyle\frac{\dr}{\dl}\bigr)$. 
\begin{lemma}\label{lemma:U_at_trivial_FP}
 \begin{align}
& U\bigl(1,\phi(1;\epsilon);\e\bigr)=1-\frac{\dr}{\dl}-\Phi(1;{\e}), \\
&U\bigl(1,\phi(1;\epsilon);\e\bigr)
\begin{cases}
 > 0\mbox{ if } \e< \e^{\mathrm{SIR}}\bigl(\textstyle\frac{\dr}{\dl}\bigr),\\
 = 0\mbox{ if } \e= \e^{\mathrm{SIR}}\bigl(\textstyle\frac{\dr}{\dl}\bigr),\\
 < 0\mbox{ if } \e> \e^{\mathrm{SIR}}\bigl(\textstyle\frac{\dr}{\dl}\bigr).
\end{cases}
\label{235136_15Jan15}
\end{align}
\end{lemma}
\noindent{\itshape Proof}: 
The first part is straightforward from \eqref{173605_11Jan15}.
The second part is obvious from the definition of SIR limit. \qed
\subsection{Potential Function at Non-Trivial Fixed Point}
\begin{figure*}[t]
\begin{picture}(250,150)(0,0)
\put(5,0){ \includegraphics[width=9cm]{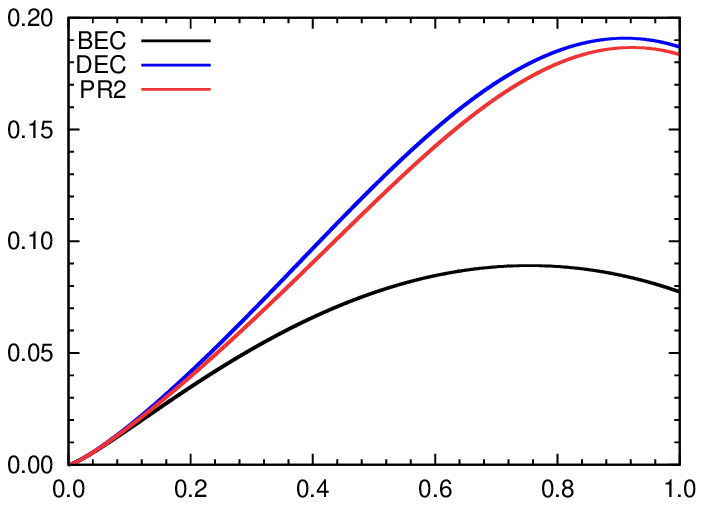} }
\put(0,50){\rotatebox{90}{$U(x_1,x_2[x_1];\epsilon[x_1])$}}
\put(142,-5){$x_1$}
\end{picture}
\begin{picture}(250,150)(0,0)
\put(5,0){ \includegraphics[width=9cm]{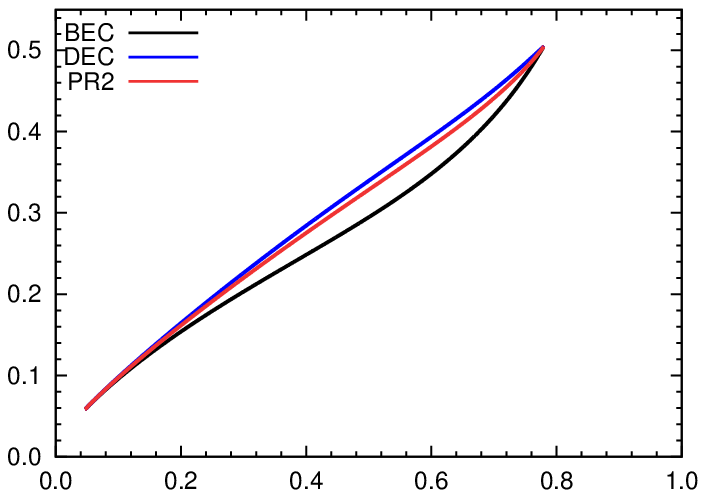} }
\put(0,50){ \rotatebox{90}{$U(x_1,x_2[x_1];\epsilon[x_1])$}}
\put(142,-5){$x_1$}
\end{picture}
\caption{Potential functions $U(x_1,x_2[x_1];\epsilon[x_1])$ at non-trivial fixed points of $(4,2,2)$ MN codes and $(6,3,3)$ MN codes for 3 types of generalized erasure channels BEC($\epsilon[x_1]$), DEC($\epsilon[x_1]$), PR2($\epsilon[x_1]$) and PR3($\epsilon[x_1]$). 
}
\label{201137_15Jan15}
\end{figure*}
Next, for given $x_1\in(0,1)$, solve \eqref{145556_10Jan15} in terms of $x_2$. Denote this by $x_2[x_1]$. 
\begin{align}
&x_2[x_1]= 1-\biggl(\frac{1-x_{1}^{\frac{1}{\dl-1}}}{(1-x_1)^{\dr-1}}\biggr)^{\frac{1}{\dg}}.
\end{align}
Note that for some $x_1$, $x_2[x_1]$ may fall outside $[0,1]$  as one can see from Fig.~\ref{201137_15Jan15}. 
 Such points are excluded from the set of fixed points. 
Then it follows that 
\begin{align}
 (x_1,x_2[x_1])\in \mathcal{F}(\epsilon[x_1]) \mbox{ iff } (x_2[x_1],\epsilon[x_1])\in [0,1]^2,\label{172507_11Jan15} 
\end{align}
 where $\epsilon[x_1]$ is the unique solution of  the equation \eqref{vasx2} with $x_2=x_2[x_1]$ holds. In precise, the equation is as follows. 
\begin{align}
x_{2}[x_1]=&\phi(\{1-(1-x_{1})^{\dr}(1-x_{2}[x_1])^{\dg-1}\}^{\dg};{\e}[x_1])\\
&\cdot\{1-(1-x_{1})^{\dr}(1-x_{2}[x_1])^{\dg-1}\}^{\dg-1}.
\end{align}
Uniqueness is due to the assumption that $\phi(x;\e)$ is strictly increasing in $\e$.
We call such fixed points \eqref{172507_11Jan15} non-trivial.

For example of BEC$(\e)$, $\e_\mathrm{BEC}[x_1]$ can be written in an explicit form. 
\begin{align}
 \e_\mathrm{BEC}[x_1]:=\frac{x_2[x_1]}{\{1-(1-x_{1})^{\dr}(1-x_{2}[x_1])^{\dg-1}\}^{\dg-1}}.\label{215219_15Jan15} 
\end{align}

Obviously, trivial and non-trivial fixed-points cover the set $\mathcal{F}(\e)$, in precise, $ \mathcal{F}(\e)=\mathcal{F}_\mathrm{t}(\e)\cup \mathcal{F}_\mathrm{n}(\e)$.
We denote the set of trivial and non-trivial fixed-points respectively by $\mathcal{F}_\mathrm{t}(\e)$ and $\mathcal{F}_\mathrm{n}(\e)$. 
\begin{align}
 \mathcal{F}_\mathrm{t}(\e)&:=\bigl\{\bigl(1,\phi(1;\e)\bigr)\bigr\}\\
\mathcal{F}_\mathrm{n}(\e)&:=\bigl\{(x_1,x_2[x_1])\mid \e[x_1]=\e, x_1\in(0,1)\bigr\}\\
\mathcal{F}_\mathrm{t}(\e)&\cap \mathcal{F}_\mathrm{n}(\e)\mbox{ is an empty set.}
\end{align}

Let $\phi[x_1]$ be the unique solution of \eqref{vasx2} in terms of $\phi(\cdot)$ for given $x_1, x_2=x_2[x_1]$ and $\e=\e[x_1]$.
 \begin{align}
\phi[x_1]:=\frac{x_2[x_1]}{\bigl(1-(1-x_{1})^{\dr}(1-x_{2}[x_1])^{\dg-1}\bigr)^{\dg-1}}.\label{214150_11Jan15}
 \end{align}
Equivalently, we have 
\begin{align}
& \phi[x_1]=\phi(\psi[x_1];\e[x_1]),\label{234603_11Jan15}
\end{align}
where
\begin{align}
\psi[x_1]:=\bigl(1-(1-x_1)^{\dr}(1-x_{2}[x_1])^{\dg-1}\bigr)^{\dg}. \label{psi}
\end{align}
Note that $x_2[x_1], \phi[x_1]$ and $\psi[x_1]$ do not depend on the channel GEC$(\phi(\cdot;\e))$,
while $\e[x_1]$ does.
From \eqref{215219_15Jan15} and \eqref{214150_11Jan15}, it can be seen that 
\begin{align}
  \e_\mathrm{BEC}[x_1]=\phi[x_1].\label{131857_15Jan15}
\end{align}

Substituting  $x_1, x_2[x_1]$ and $\phi[x_1]$ into \eqref{173605_11Jan15},
at non-trivial fixed points $(x_1, x_2[x_1])\in \mathcal{F}(\epsilon[x_1])$, 
we have the potential function  as
\begin{align}
&U(x_1, x_2[x_1];\e[x_1])\label{133030_15Jan15}\\
&=1 -\Phi(\psi[x_1];{\epsilon[x_1]})  \\
&- \frac{\dr}{\dl}\{1-(1-x_1)^{\dr-1}(1-x_2[x_1])^{\dg}\}^{\dl}  \\
&- (1-x_1)^{\dr}(1-x_2[x_1])^{\dg} \cdot \Bigl(1+\frac{\dr x_1}{1-x_1}+\frac{\dg x_2[x_1]}{1-x_2[x_1]}\Bigr).
\end{align}

From \cite[Lemma 4]{ISIT_OJKP} and \cite[Lemma 1, Lemma2]{6874884}, we have the following lemma.
\begin{lemma}\label{lemma:BEC>0}
Consider transmissions over {\rm BEC}($\epsilon$), i.e., we have $\phi(x)=\phi_{\mathrm{BEC}}(x)=\epsilon$ and 
$\phi[x_1]=\e[x_1]$.
Let $U_{\mathrm{BEC}}(x_1, x_2[x_1];\e_{\mathrm{BEC}}[x_1])$ be
the potential function  of $(\dl,\dr,\dg)$  MN codes at the non-trivial fixed points. 
For any $\dl>\dr$ it holds that 
\begin{align}
U_{\mathrm{BEC}}(x_1, x_2[x_1];{\e}_\mathrm{BEC}[x_1]) > 0   \mbox{ for }   x_1 \in (0, 1) \label{eq:inU}
\end{align}
for $(\dr,\dg)=(2,2)$ and $(3,3)$. 
\end{lemma}

\begin{lemma}\label{GEC>=BEC}
For any {\rm GEC}($\phi(x;\e)$), let $U(x_1, x_2[x_1];\e[x_1])$ be the potential function at the non-trivial fixed point 
 $(x_1,x_2[x_1])\in \mathcal{F}(\epsilon[x_1])$ for $x_1 \in (0, 1)$ 
as defined in \eqref{133030_15Jan15}. 
Then, it holds that
\begin{align}
U(x_1, x_2[x_1];\e[x_1]) \ge &U_{\mathrm{BEC}}(x_1,x_2[x_1];\e_\mathrm{BEC}[x_1]). \label{eq:gecbecu}
\end{align}
\end{lemma}
\noindent {\itshape Proof}: From \eqref{172507_11Jan15}, we have $(x_2[x_1],\e[x_1])\in [0,1]^2$. From \eqref{psi}, we have $\psi[x_1]\in[0,1]$. 
From this and using  \eqref{234603_11Jan15} and \eqref{131857_15Jan15}, we obtain $\e_{\mathrm{BEC}}[x_1]\in [0,1]$.
This justifies that $U_{\mathrm{BEC}}(x_1,x_2[x_1];\e_\mathrm{BEC}[x_1])$ is well-defined since $x_1,x_2[x_1],\e_\mathrm{BEC}[x_1]\in[0,1]$. We have
\begin{align}
U(&x_1, x_2[x_1];\e[x_1]) -U_{\mathrm{BEC}}(x_1,x_2[x_1];\e_\mathrm{BEC}[x_1])\\
 \stackrel{\mathrm{(a)}}{=}&\e_{\mathrm{BEC}}[x_1]\cdot(1-(1-x_1)^{\dr}(1-x_2[x_1])^{\dg-1})^{\dg} \\
  &-\Phi\bigl((1-(1-x_1)^{\dr}(1-x_2[x_1])^{\dg-1})^{\dg};\e[x_1]\bigr) \\
 \stackrel{\mathrm{(b)}}{=}& \phi[x_1]\cdot\psi[x_1] - \Phi\bigl(\psi[x_1];\e[x_1]\bigr) \\
  =& \phi[x_1]\cdot\psi[x_1] - \int_0^{\psi[x_1]} \phi(x';\e[x_1])dx' \\
  \stackrel{\mathrm{(c)}}{=}& \int_0^{\psi[x_1]} \phi[x_1] -  \phi(x';\e[x_1])dx' \\
 \stackrel{\mathrm{(d)}}{=}& \int_0^{\psi[x_1]} \phi(\psi[x_1];\e[x_1]) -  \phi(x';\e[x_1])dx' \\
 \stackrel{\mathrm{(e)}}{\ge}& 0.
\end{align}
The equality (a)  is due to \eqref{133030_15Jan15} and \eqref{133138_15Jan15}.
The equality (b)  is due to \eqref{131857_15Jan15}.
In (c), we used the fact that $\psi[x_1]$ does not depend on channel GEC$(\phi(\cdot;\e))$. 
The equality (d)  is due to \eqref{234603_11Jan15}.
In (e), we used the fact that $\phi(\psi[x_1];\e[x_1])\ge \phi(x';\e[x_1])$ for $x'\in \bigl[0,\psi[x_1]\bigr]$ since $\phi(x;\e)$ is non-decreasing in $x$.
The equality is attained with $\phi(x;\e)=\phi_\mathrm{BEC}(x;\e)=\e$. \qed

 In Fig.~\ref{201137_15Jan15}, we plotted the potential functions at non-trivial fixed points of $(4,2,2)$ MN codes and $(6,3,3)$ MN codes for 3 types of generalized erasure channels BEC($\epsilon[x_1]$), DEC($\epsilon[x_1]$) and PR2($\epsilon[x_1]$). 
 One can see that the potential function for BEC is lower than other curves for any $x_1\in(0,1)$ as claimed in Lemma \ref{GEC>=BEC}. 

\subsection{Potential Threshold Equals to SIR Limit}
Next theorem shows 
the potential threshold  of some MN codes is equal to the SIR limit. 
\begin{theorem}
For any GEC($\phi(x;\epsilon)$), the potential threshold $\epsilon^*$ of $(\dl,\dr,\dg)$ MN codes is equal to $\e^{\mathrm{SIR}}(\frac{\dr}{\dl})$ 
for any $\dl>\dr$, $(\dr,\dg)=(2,2)$ and $(3,3)$.
\end{theorem}
\noindent{\itshape Proof}: 
From Lemma \ref{lemma:BEC>0} and Lemma \ref{GEC>=BEC},
we have that for any
 $x_1\in(0,1) $ such that 
$\e=\e[x_1]\in \bigl[0,\e^{\mathrm{SIR}}(\textstyle{\frac{\dr}{\dl}})\bigr),$
\begin{align}
&U(x_1,x_2[x_1];\e[x_1])>0. \label{003454_16Jan15} 
\end{align}
It follows that 
\begin{align}
&     \e^{*} = \sup \{ \e \in [0,1] \mid\min_{(x_1,x_2) \in \mathcal{F}(\e)} U(x_1,x_2;\e) > 0 \}\\
    &=\sup \{ \e \in [0,1] \mid\displaystyle{\min_{(x_1,x_2)\in \mathcal{F}_\mathrm{t}(\e)\cup \mathcal{F}_\mathrm{n}(\e)}} U(x_1,x_2;\e) > 0 \}\\
    &\stackrel{\mathrm{(a)}}{=}\sup \{ \e \in \bigl[0,\e^{\mathrm{SIR}}(\textstyle{\frac{\dr}{\dl}})\bigr) \mid\displaystyle{\min_{(x_1,x_2)\in \mathcal{F}_\mathrm{t}(\e)\cup \mathcal{F}_\mathrm{n}(\e)}} U(x_1,x_2;\e) > 0 \}\\
    &\stackrel{\mathrm{(b)}}{=}\e^{\mathrm{SIR}}\bigl(\textstyle{\frac{\dr}{\dl}}\bigr).
\end{align}
In (a), we used \eqref{235136_15Jan15}.
In (b), we used \eqref{235136_15Jan15} and \eqref{003454_16Jan15}. \qed

 Define the BP threshold $\e^{\mathrm{BP}}(\dl,\dr,dg,L,w)$ as the spremum value of $\e$ such that 
 the SC system with chain length $L$ and coupling width $w$ of $(\dl,\dr,\dg)$ MN codes over GEC$(\e)$ converges to zero. 
In precise, $\e^{\mathrm{BP}}(\dl,\dr,dg,L,w)$ is 
\begin{align}
 \sup\{\e\in[0,1]|\lim_{t\to\infty}x_i^{(t)}=0 \allowbreak \mbox{ for }i=0,1,\dotsc,L-1\}.
\end{align}
Then, from \eqref{010653_25Jan15} and Theorem \ref{thorem:potential threshold = SIR limit} and \ref{theorem:sc_theorem} we have the following theorem.
\begin{theorem}
\label{thorem:potential threshold = SIR limit}
For any GEC($\phi(x;\epsilon)$), the potential threshold $\epsilon^*$ of $(\dl,\dr,\dg)$ MN codes is equal to $\e^{\mathrm{SIR}}(\frac{\dr}{\dl})$ 
for any $\dl>\dr$, $(\dr,\dg)=(2,2)$ and $(3,3)$.
\begin{align}
& \lim_{w\to\infty}\lim_{L\to\infty}\e^{\mathrm{BP}}(\dl,\dr,dg,L,w)=\e^{\mathrm{SIR}}(\dr/\dl),\\
&\lim_{w\to\infty}\lim_{L\to\infty}R^{\rm{MN}}(\dl,\dr,\dg,L,w)=\frac{\dr}{\dl}.
\end{align}
In words, some SC-MN codes universally achieve the SIR limit of any GEC($\phi(x;\epsilon)$) in the limit of large $L$ and $w$. 
\end{theorem}
 \section{Conclusion and Future Work}
We have shown that  some SC-MN codes achieve the SIR limit of any GEC($\phi(x;\epsilon)$) via potential function. 
The future works include an extension erasure multi-acess channels \cite{KuKaMAC} and to more general channels, e.g., PR2 channels with Gaussian noise.

\bibliographystyle{IEEEtran}
\bibliography{IEEEabrv,kenta_bib} 

\appendix
\begin{figure*}
\begin{picture}(250,150)(0,0)
\put(5,0){ \includegraphics[width=9cm]{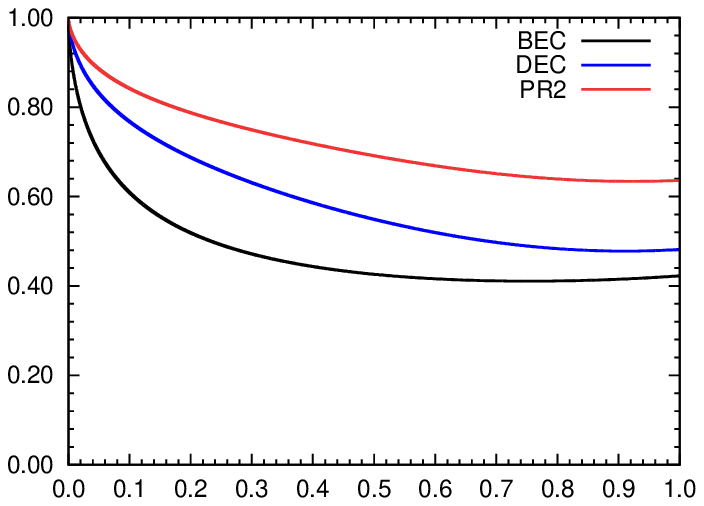} }
\put(0,85){\rotatebox{90}{$\epsilon[x_1]$}}
\put(142,-5){$x_1$}
\end{picture}
\begin{picture}(250,150)(0,0)
\put(5,0){ \includegraphics[width=9cm]{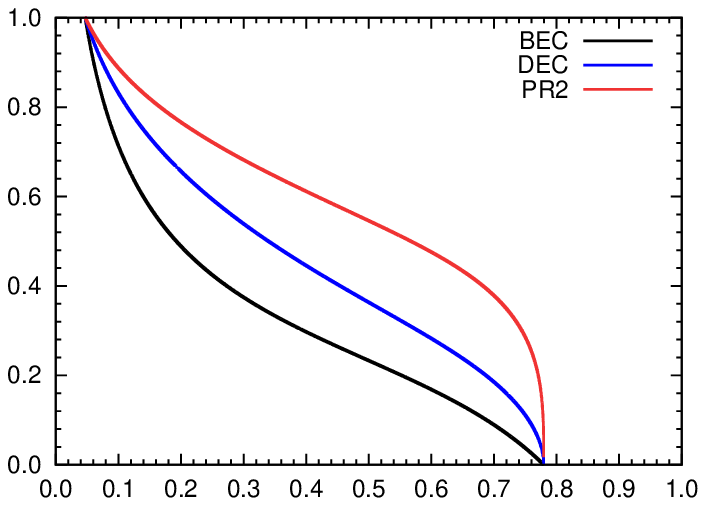} }
\put(0,85){ \rotatebox{90}{$\epsilon[x_1]$}}
\put(142,-5){$x_1$}
\end{picture}
\caption{Plot of $\epsilon[x_1]$ at non-trivial fixed points of $(4,2,2)$ MN codes and $(6,3,3)$ MN codes for 3 types of generalized erasure channels BEC($\epsilon[x_1]$), DEC($\epsilon[x_1]$), 
PR2($\epsilon[x_1]$) and 
PR3($\epsilon[x_1]$). 
}
\label{201137_15Jan15}
\end{figure*}
In Appendix, we give three GECs  as example. 
\begin{example}[Binary Erasure Channel]
For the binary erasure channel BEC($\e$) with erasure probability $\epsilon$, 
 $\phi(x;\epsilon), \Phi(x;\epsilon), I(\epsilon)$ and  $\e[x_1]$ are respectively given as 
\begin{align}
 \phi_{\mathrm{BEC}}(x;\epsilon)&=\epsilon,\\
 \Phi_{\mathrm{BEC}}(x;\epsilon)&=\epsilon x,\\
 I_\mathrm{BEC}(\epsilon)&=1-\epsilon,\\
 \e[x_1]&=\e_\mathrm{BEC}[x_1], 
\end{align}
where $\e_\mathrm{BEC}[x_1]$ is the unique solution $\e_\mathrm{BEC}$ of the following equation. 
\begin{align}
 x_{2}[x_1]=&\phi_\mathrm{BEC}(\{1-(1-x_{1})^{\dr}(1-x_{2}[x_1])^{\dg-1}\}^{\dg};\e_\mathrm{BEC}[x_1])\\
 &\cdot\{1-(1-x_{1})^{\dr}(1-x_{2}[x_1])^{\dg-1}\}^{\dg-1}. 
\end{align}
$\e_\mathrm{BEC}[x_1]$ can be written in an explicit form \eqref{215219_15Jan15}.

The potential function is given as 
\begin{align}
U_\mathrm{BEC}&(x_1,x_2[x_1];\e[x_1])\label{133138_15Jan15} \\
:=1 &-\e_\mathrm{BEC}[x_1]\cdot(\{1-(1-x_1)^{\dr}(1-x_2[x_1])^{\dg-1}\}^{\dg})  \\
&- \frac{\dr}{\dl}\{1-(1-x_1)^{\dr-1}(1-x_2[x_1])^{\dg}\}^{\dl}  \\
&- (1-x_1)^{\dr}(1-x_2[x_1])^{\dg} \\
&\cdot \Bigl(1+\frac{\dr x_1}{1-x_1}+\frac{\dg x_2[x_1]}{1-x_2[x_1]}\Bigr).
\end{align}
\end{example}

\begin{example}[Dicode Erasure Channel]
 For the dicode erasure channel DEC($\e$) with erasure probability $\epsilon$, 
the output $\underline{y}\in \{0,1,-1,?\}^n$ for given $\underline{x}\in \{0,1\}^n$ is defined as follows. 
\begin{align}
 y_j= 
 \begin{cases}
  x_j -x_{j-1}& \mbox{w.p. } 1-\epsilon,\\
  ? & \mbox{w.p. } \epsilon,
 \end{cases}
\end{align}
where $x_0:=0$. 
 $\phi(x;\epsilon), \Phi(x;\epsilon), I(\epsilon)$ and  $\e[x_1]$ are respectively given as 
\begin{align}
 \phi_{\mathrm{DEC}}(x;\epsilon)&=\frac{4\epsilon^2}{(2-x(1-\epsilon))^2}\\
 \Phi_{\mathrm{DEC}}(x;\epsilon)&=\frac{4\e^2}{(1-\e)(2-(1-\e)x)},\\
 I_\mathrm{DEC}(\epsilon)&=1-\frac{2\epsilon^2}{1+\epsilon},\\
 \e[x_1]&=\e_\mathrm{DEC}[x_1],
\end{align}
where $\e_\mathrm{DEC}[x_1]$ is the unique solution $\e_\mathrm{DEC}$ of the following equation. 
\begin{align}
 x_{2}[x_1]=&\phi_\mathrm{DEC}(\{1-(1-x_{1})^{\dr}(1-x_{2}[x_1])^{\dg-1}\}^{\dg};\e_\mathrm{DEC})\\
 &\cdot\{1-(1-x_{1})^{\dr}(1-x_{2}[x_1])^{\dg-1}\}^{\dg-1}. 
\end{align}
In the case of {\rm DEC}($\epsilon$), $\e_\mathrm{DEC}[x_1]$ can be written in an explicit form. 
\begin{align}
 \e_\mathrm{DEC}[x_1]:=\frac{(2-\psi[x_1])\Bigl(\frac{x_2[x_1]}{\psi[x_1]^{{(\dg-1)}/{\dg}}}\Bigr)^{1/2}}{2-\psi[x_1]\Bigl(\frac{x_2[x_1]}{\psi[x_1]^{{(\dg-1)}/{\dg}}}\Bigr)^{1/2}}
\end{align}
In general, we do not need the explicit form of $\e[x_1]$ for the purpose of this paper.
The potential function is given as follows. 
\begin{align}
\label{132949_15Jan15} 
U_\mathrm{DEC}&(x_1,x_2[x_1];\e[x_1]) \\
:=1 &-\Phi_\mathrm{DEC}(\{1-(1-x_1)^{\dr}(1-x_2[x_1])^{\dg-1}\}^{\dg};\e_\mathrm{DEC}[x_1]) \\
&- \frac{\dr}{\dl}\{1-(1-x_1)^{\dr-1}(1-x_2[x_1])^{\dg}\}^{\dl}  \\
&- (1-x_1)^{\dr}(1-x_2[x_1])^{\dg} \\
&\cdot \Bigl(1+\frac{\dr x_1}{1-x_1}+\frac{\dg x_2[x_1]}{1-x_2[x_1]}\Bigr).
\end{align}
\end{example}
\begin{example}[Partial Response 2 Channel]
 For the PR2 channel PR2($\e$) with erasure probability $\epsilon$, 
the output $\underline{y}\in \{0,1,2,3,4,?\}^n$ for given $\underline{x}\in \{0,1\}^n$ is defined as follows. 
\begin{align}
 y_j= 
 \begin{cases}
  x_j +2x_{j-1}+x_{j-2}& \mbox{w.p. } 1-\epsilon,\\
  ? & \mbox{w.p. } \epsilon,
 \end{cases}
\end{align}
where $x_0:=0$. 
 $\phi(x;\epsilon), \Phi(x;\epsilon), I(\epsilon)$ and  $\e[x_1]$ are respectively given as 
\begin{align}
 \phi_{\mathrm{PR2}}(x;\epsilon)&=\frac{4\e^3\bigl(4-4(1-\e)x+(1-\e)x^2\bigr)}{(4-2(1-\e^2)x-(1-\e)\e^2 x^2)^2}\\
 \Phi_{\mathrm{PR2}}(x;\epsilon)&=
4\e\Bigl(\frac{1}{2} - \frac{2 - x}{4 - 2(1 - \e^2)x - (1 - \e)\e^2x^2}\Bigr),\\
 I_\mathrm{PR2}(\epsilon)&=1-\frac{2\epsilon^3(\e+1)}{\e^3+\e_2+2},\\
 \e[x_1]&=\e_\mathrm{PR2}[x_1],
\end{align}
where $\e_\mathrm{PR2}[x_1]$ is the unique solution $\e_\mathrm{PR2}$ of the following equation. 
\begin{align}
 x_{2}[x_1]=&\phi_\mathrm{PR2}(\{1-(1-x_{1})^{\dr}(1-x_{2}[x_1])^{\dg-1}\}^{\dg};\e_\mathrm{PR2})\\
 &\cdot\{1-(1-x_{1})^{\dr}(1-x_{2}[x_1])^{\dg-1}\}^{\dg-1}. 
\end{align}
The potential function is given as follows. 
\begin{align}
U_\mathrm{PR2}&(x_1,x_2[x_1];\e[x_1]) \\
:=1 &-\Phi_\mathrm{PR2}(\{1-(1-x_1)^{\dr}(1-x_2[x_1])^{\dg-1}\}^{\dg};\e_\mathrm{PR2}[x_1]) \\
&- \frac{\dr}{\dl}\{1-(1-x_1)^{\dr-1}(1-x_2[x_1])^{\dg}\}^{\dl}  \\
&- (1-x_1)^{\dr}(1-x_2[x_1])^{\dg} \\
&\cdot \Bigl(1+\frac{\dr x_1}{1-x_1}+\frac{\dg x_2[x_1]}{1-x_2[x_1]}\Bigr).
\end{align}
\end{example}
\end{document}